\documentclass[12pt]{iopart}

\usepackage{textcomp} 
\usepackage{iopams}
\usepackage{graphicx}
\usepackage{wasysym}
\usepackage{amssymb,amsfonts}
\usepackage{url}   

\begin{document}

\title[Differential atom interferometry with $^{87}$Rb and $^{85}$Rb]{Differential atom interferometry with $^{87}$Rb and $^{85}$Rb for testing the UFF in STE-QUEST}

\author{C~Schubert$^1$, J~Hartwig$^1$, H~Ahlers$^1$, K~Posso-Trujillo$^1$, N~Gaaloul$^1$, U~Velte$^1$,
A~Landragin$^2$, A~Bertoldi$^3$, B~Battelier$^3$, P~Bouyer$^3$,
F~Sorrentino$^4$, G~M~Tino$^4$,
M~Krutzik$^5$, A~Peters$^5$, S~Herrmann$^6$, C~L\"ammerzahl$^6$, L~Cacciapouti$^8$, E~Rocco$^7$, K~Bongs$^7$, W~Ertmer$^1$ and E~M~Rasel$^1$}

\address{
$^1$Institute of Quantum Optics, Leibniz University Hanover,
Welfengarten 1, 30167 Hanover, Germany\\
$^2$LNE-SYRTE, Observatoire de Paris, CNRS and UPMC, 61 avenue de l'observatoire 75014 Paris, France\\
$^3$Laboratoire Photonique, Num{\'e}rique et Nanosciences--LP2N Universit{\'e} Bordeaux--IOGS--CNRS: UMR 5298, Talence, France\\
$^4$Dipartimento di Fisica e Astronomia and LENS Laboratory, Universit{\`a} di Firenze - INFN, Sezione di Firenze - via G. Sansone 1, 50019 Sesto Fiorentino (Firenze), Italy\\
$^5$Humboldt-Universität zu Berlin, Institut für Physik, Newtonstr. 15, 12489 Berlin,
Germany\\
$^6$Center of Applied Space Technology and Microgravity (ZARM), University
Bremen, Am Fallturm, 28359 Bremen, Germany\\
$^7$School of Physics and Astronomy, University of Birmingham, Birmingham, B15
2TT, United Kingdom\\
$^8$ESA - European Space Agency, ESTEC, Keplerlaan 1, 2200 AG Noordwijk ZH,
Netherlands}
\ead{schubert@iqo.uni-hannover.de}

\begin{abstract}
In this paper we discuss in detail an experimental scheme to test the universality of free fall (UFF) with a differential $^{87}$Rb / $^{85}$Rb atom interferometer applicable for extended free fall of several seconds in the frame of the STE-QUEST mission. This analysis focuses on suppression of noise and error sources which would limit the accuracy of a violation measurement. We show that the choice of atomic species and the correctly matched parameters of the interferometer sequence are of utmost importance to suppress leading order phase shifts. In conclusion we will show the expected performance of $2$ parts in $10^{15}$ of such an interferometer for a test of the UFF.
\end{abstract}

\maketitle
\section{Introduction}
The singular importance of the theory of relativity and in this respect the universality of free fall (UFF) motivates precision tests of composition dependent couplings with gravity at the frontier of measurement technology. Dedicated experiments like lunar laser ranging~\cite{Williams2004PRL} and torsion balances~\cite{Schlamminger2008PRL} exclude violations of the universality of free fall by testing the E\"otv\"os ratio above a level of $\eta=10^{-13}$, but are still just scraping the edges of predicted violation scenarios. While a new generation of classical experiments~\cite{Touboul12CQG,Nobili12CQG} plans to utilize space borne platforms to decrease the influence of disturbances as experienced on ground and the well developed measurements are continuously improving, atom interferometry opens new pathways to test the UFF~\cite{Fray2004PRL,Bonnin2013PRL}. Atom interferometers allow the investigation of formerly inaccessible atomic species, tests with absolute isotopic purity of the samples, and the use of specific quantum mechanical states of these samples~\cite{Fray2004PRL}. Recent publications even propose the use of antimatter for a UFF test, certainly not possible with classical instruments~\cite{Mueller2013arXiv}. Following these developments, STE-QUEST (Spacetime Explorer and Quantum Equivalence Principle Space Test)~\cite{missionpaper2013,ESA_STEQUEST13website}, a medium size (M3) satellite mission proposed to ESA in the scope of the Cosmic Vision program, aims for reaching an uncertainty in the E\"otv\"os ratio of $2$ parts in $10^{15}$ with a dual species $^{87}$Rb/$^{85}$Rb atom interferometer. Previously, a preliminary status of the atom interferometer payload was published in~\cite{Tino2013NPB}.\\
The capabilities of atom interferometry for testing the UFF were first shown by a comparison of a Cs interferometer with a laser gravimeter using a macroscopic test mass at a level of $ 7\cdot10^{-9}$~\cite{Peters1999Nat}. A comparison between $^{87}$Rb and $^{85}$Rb atom interferometers was later demonstrated at a level of $10^{-7}$~\cite{Fray2004PRL,Bonnin2013PRL}. In the wake of these demonstrations a new class of dedicated UFF-testing atom interferometry experiments are in the process of being implemented~\cite{Dickerson2013PRL,Dimopoulos2007PRL}, aiming for measurement accuracies comparable to the state of the art classical experiments and complementing them for a complete coverage of possible sample parameters~\cite{Hohensee2011JMO}. At the same time, atom interferometry in zero-g environments was demonstrated to investigate the unique conditions which allow extended free evolution times in the regime of several seconds, shallower traps compared to Earth-based setups, and the technology readiness for space borne operation~\cite{Nyman2006APB,Zoest2010Science,Geiger2011NatComm,Muntinga2013PRL} as in STE-QUEST. \\
Atom interferometers exploit the wave nature of massive particles in a role reversal between matter and light. Beam splitters made of light allow for coherent manipulation of matter waves to engineer geometries analogous to Mach-Zehnder interferometers in optics. During each atom-light interaction, the position of the atoms is referenced with respect to the beam splitting light field. A relative acceleration between interactions changes the phase in the atom interferometer output ports. To define the beam splitting light field, a retroreflected setup is used~\cite{Peters1999Nat}. Thus, the retroreflection mirror is effectively the reference. For the UFF test, two atom interferometers with different species operate simultaneously with the same reference mirror. This enables a common mode rejection of systematics and noise sources, most prominently vibrations. The achievable suppression ratio improves for similar atomic species~\cite{Bonnin2013PRL,Varoquaux2009NJP}. Consequently, a trade-off between the expected UFF-violating signal, increased for very different atomic species~\cite{Damour2012CQG,Hohensee2011JMO}, and the suppression factor for environmental noise achievable for the two atomic species is necessary.\\
In this publication, we will motivate and discuss the very beneficial situation with the choice of two isotopes of the same element, namely $^{87}$Rb and $^{85}$Rb, for STE-QUEST. We will present the measurement scheme based on two atomic species in a microgravity environment utilizing a double diffraction atom interferometer. Then, we will explain the leading order phase contributions perturbing the science signal and discuss the expected influence of external parameters on the measurement contrast. Finally, we estimate the STE-QUEST sensitivity to UFF tests taking into account the specified performance of the instrument and the spacecraft orbit.
\section{Orbit}
\label{sec:orbit}
\begin{figure}[tp]
\centering
\includegraphics*[width=0.65\textwidth]{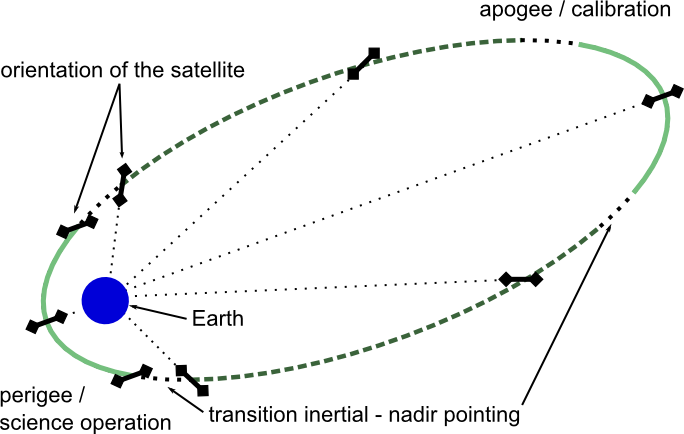}
\caption{Reference orbit for the STE-QUEST satellite. During perigee and apogee pass the spacecraft is in an inertial pointing mode. In between a nadir pointing mode enables the clock comparison measurements.}
\label{fig:orbit}
\end{figure}
The orbit for STE-QUEST~\cite{ESA_STEQUEST13website,SciRD2013,Yellowbook2013} has to support both the atom interferometer measurements for testing the UFF and clock comparisons for gravitational redshift tests. The orientation of the atom interferometer sensitive axis along the STE-QUEST orbit is shown in figure~\ref{fig:orbit} and orbit parameters are listed in table~\ref{tab:orbit_parameters}. Requirements from the two objectives are complementary: The redshift tests need a high distance to Earth to enable long common view contacts between ground stations, while the UFF test requires proximity to Earth. A highly elliptical orbit was chosen as a trade-off. Consequently, UFF science data is taken during perigee pass when the spacecraft realizes an inertial pointing mode as depicted in figure~\ref{fig:orbit}. Remaining time is used by the atom interferometer for calibration and verification measurements. Especially around apogee, it is operated to generate a null signal as the local gravitational acceleration is negligible. This is beneficial for the suppression of error terms that are stable in time. Detailed environmental parameters are listed in table~\ref{tab:orbit_requirements}.
 \begin{table}[tb]
 \begin{center}
 \caption{Parameters for the STE-QUEST reference orbit
 .}
 \begin{tabular}{p{0.35\textwidth}p{0.25\textwidth}p{0.3\textwidth}}\hline
   \textbf{Parameter} & \textbf{Perigee} & \textbf{Apogee} \\ \hline
  Altitude in $\mathrm{km}$ & $700$-$2200$ & $51000$ \\
  Gravitational acceleration in $\mathrm{ms}^{-2}$ & $8$-$5.4$ & $0.12$ \\ 
  Angular velocity in $\mathrm{rads}^{-1}$ & $1.4\cdot10^{-3}$ & $2.2\cdot10^{-5}$ \\ 
  Grav. gradient in $\mathrm{s}^{-2}$ & $2.2\cdot10^{-6}$ & $4.2\cdot10^{-9}$ \\ 
  Drag in $\mathrm{ms}^{-2}$ & $<10^{-6}$ & \\ \hline
 \end{tabular}
 \label{tab:orbit_parameters}
 \end{center}
 \end{table}
 
 \begin{table}[tb]
 \begin{center}
 \caption{Detailed constraints for the STE-QUEST atom interferometer
 .}
 \begin{tabular}{p{0.35\textwidth}p{0.25\textwidth}p{0.3\textwidth}}\hline 
  \textbf{Parameter} & \textbf{Value} & \textbf{Comment} \\ \hline
  $\Delta a$ sensitivity in $\mathrm{ms}^{-2}\mathrm{Hz}^{-1/2}$ & $1.3\cdot10^{-11}$ &  \\ 
  $\eta$ inaccuracy & $2\cdot10^{-15}$ &  \\ \hline
  Local grav. acc. in $\mathrm{ms}^{-2}$ & $>3$ & During Science op. \\ \hline
  Gravity gradient in $\mathrm{s}^{-2}$ & $<2.5\cdot10^{-6}$ &  \\ \hline
    Spacecraft self-gravity & below Earth's contribution at perigee &  \\ \hline
  Non grav. acc. in $\mathrm{ms}^{-2}$ & $<4\cdot10^{-7}$  & Along sensitive axis  \\
   & $<1\cdot10^{-6}$ & Orthogonal axes  \\      - PSD in $\mathrm{ms}^{-2}\mathrm{Hz}^{-1/2}$ at frequency $\nu$ in $\mathrm{Hz}$ & $10^{-3}\cdot\nu\,\mathrm{Hz}^{-1}$;  $2\cdot10^{-5}$ at $[0.001,0.02]$; $[0.02,100]$ &  \\ 
  - rms in $\mathrm{ms}^{-2}$ at frequency $\nu$ in $\mathrm{Hz}$ & $<4\cdot10^{-7}$;$4\cdot10^{-5}\cdot\nu\,\mathrm{Hz}^{-1}$; $4\cdot10^{-4}$ at $[0,0.01]$; $[0.01,10]$; $>10$ &  \\ \hline
  Nadir pointing at perigee & Better than $3\,^{\circ}$ &  \\
  Rotations & $10^{-6}\,\mathrm{rads}^{-1}$ & Uncertainty $10^{-7}\,\mathrm{rads}^{-1}$  \\ \hline
  Magnetic fields in \textmu$\mathrm{G}$ & $<\pm10$ & Inside magnetic shield \\ 
  - gradients in \textmu$\mathrm{Gm}^{-1}$ & $<\pm4$ & Inside magnetic shield \\ 
  - variations in $\mathrm{G}$ at $\nu$ in $\mathrm{Hz}$ & $1$; $0.1$; $0.01$; $0.1$; $1$ at $[0,0.001]$; $[0.01]$; $[0.1,10]$; $100$; $1000$ & Outside magnetic shield; suppressed by $10^{4}$ \\ \hline
  Revolution in $\mathrm{h}$ & $16$ & \\
  - science operation & $0.5$ & \\
  - calibration & min. $7.5$ & \\ \hline
 \end{tabular}
 \label{tab:orbit_requirements}
 \end{center}
 \end{table}
\section{Measurement Principle}
\label{sec:measurement_principle}
The measurement principle is based on a simultaneous acceleration measurement $a_{87}$ for $^{87}$Rb and $a_{85}$ for $^{85}$Rb from which the differential acceleration $\Delta a=a_{87}-a_{85}$ is extracted. By relating the differential acceleration to the projection of local $g$ in direction of the sensitive axis the E\"otv\"os ratio $\eta=|\Delta a|/g$ is derived. \\
The interferometer is implemented by double diffraction beam splitters~\cite{Leveque2009PRL,Giese2013PRA} in a Mach-Zehnder-like $\pi/2$-$\pi$-$\pi/2$ configuration. Simultaneously applied, the beam splitters manipulate a $^{87}$Rb and a $^{85}$Rb Bose-Einstein-condensate (BEC) acting as symmetric splitters in two states with opposite momentum, mirrors, and recombiners, see figure~\ref{fig:int_sequence}. The population of the output ports corresponds to the interferometer signal and is read out via fluorescence detection. Key feature is a high correlation between both interferometers achieved via careful parameter adjustment. 

\subsection{Motivation for the scheme}
The atom interferometer has to be operated under the conditions specified in section~\ref{sec:orbit}, with a launch date around $2022$. In addition, tests in the environment expected along the STE-QUEST orbit are expected to be successfully passed by $2015$. Thus, the engineering of sufficient suppression ratios for noise and bias terms and the technology readiness level of the instrument drive the choice of species for the STE-QUEST interferometer. Since the differential phase corresponding to a violation of the UFF has to be recovered from the vibrational background, a high suppression of spurious accelerations is necessary. The choice of $^{87}$Rb and $^{85}$Rb atoms is suited to this requirement. By adjusting the beam splitter laser frequencies and intensities the effective wave numbers can be matched to $10^{-9}$ and the Rabi frequencies to $10^{-4}$. Additionally, both beam splitters are switched with the same optical element matching the pulse timings. In this way, a theoretical suppression ratio for vibrations and spurious accelerations of $2.5\cdot 10^{-9}$ can be achieved, significantly reducing the statistical and systematic errors. A dual species atom interferometer with $^{87}$Rb and $^{85}$Rb relies on readily available and well tested technology~\cite{Muntinga2013PRL,Louchet2011NJP,Altin2010RSI}, indeed this combination is the only one already used in atom interferometer UFF tests~\cite{Fray2004PRL,Bonnin2013PRL}. Other atom interferometry experiments will compare $^{87}$Rb to a K isotope~\cite{Nyman2006APB,Rudolph2011MST,IQ13website}. However, as discussed in~\cite{Varoquaux2009NJP} the suppression ratio was estimated to $\sim 10^{-2}$ for the two species, limited by the different wave numbers which cannot be matched by state-of-the-art technology. A test with $^{6}$Li / $^{7}$Li is proposed in~\cite{Hohensee2011JMO}, but compared to $^{87}$Rb / $^{85}$Rb or $^{87}$Rb / K the maturity of the required technology is significantly less advanced. Considering possible violation strengths following~\cite{Damour2012CQG,Hohensee2011JMO}, $^{87}$Rb / $^{85}$Rb would lack behind $^{87}$Rb / K, and $^{6}$Li / $^{7}$Li by about two orders of magnitude. On the contrary, the suppression ratio for common-mode acceleration noise and the technical maturity for a $^{87}$Rb / $^{85}$Rb interferometer are significantly higher, overcoming this drawback.\\ 
Using BECs has the advantage of a low expansion rate which leads to atomic ensembles with a small diameter even after seconds of free evolution. Ultra-cold samples are therefore essential for achieving a high signal to noise ratio at detection and for controlling systematic effects on the measurement. With the anticipated velocity spread of $82\,$\textmu$\mathrm{m/s}$, no degradation due to the velocity selectivity of the beam splitting process is expected. Dephasing over the atomic ensemble because of velocity dependent phase shifts might still lead to moderate reduction in contrast to about $60\,\%$ (section~\ref{sec:Interferometer_contrast}). Effects of residual mean field energy on statistical and systematic errors, as discussed in section~\ref{Mean field}, become negligible.\\
The double diffraction scheme leads to a symmetric splitting of the interferometer arms, which inherently suppresses certain noise and bias terms.
\begin{table}[t]
\begin{center}
\caption{Key parameters. The sensitivity for differential accelerations which is integrated throughout the mission (see section~\ref{sec:performance}) is the top level requirement. All other parameters were deduced. Considering a Mach-Zehnder-like $\pi/2$-$\pi$-$\pi/2$ interferometer configuration, the free evolution time specifies the interval between two consecutive beam splitter pulses. The sensitivity to differential accelerations is reached under the assumptions of a contrast $C=0.6$ (section~\ref{sec:Interferometer_contrast}) and shot noise limited measurement of the differential phase. It corresponds to $2.93\cdot10^{-12}\,\mathrm{ms}^{-2}$ in $20\,\mathrm{s}$. The anticipated cycle time include preparation of the BECs, the interferometer pulse sequence, and detection. BEC expansion rate and radius correspond to the
 $1/e$ radius of a Gaussian distribution fit. }
\begin{tabular}[top]{llcc}
\hline
 Parameter & Variable & Value \\ \hline
 Sensitivity for & $\sigma_{\Delta\mathrm{a}}$ & $1.3\cdot 10^{-11}$ \\ 
 differential accelerations &  & $\mathrm{ms}^{-2}\mathrm{Hz}^{-1}$ \\ \hline
 $^{87}$Rb and $^{85}$Rb atom number & $N_{87}=N_{85}=N$ & $10^{6}$ \\
 Effective wave number & $k$ & $8\pi/(780\,\mathrm{nm})$ \\
 Free evolution time & $T$ & $5\,\mathrm{s}$ \\
 Cycle time & $T_{\mathrm{c}}$ & $20\,\mathrm{s}$ \\
 BEC expansion rate & $\sigma_{\mathrm{v}}$ & $82\,$\textmu$\mathrm{m/s}$ \\
 BEC radius & $\sigma_{\mathrm{r}}$ & $30\,$\textmu$\mathrm{m}$ \\
\hline
\end{tabular}
\label{Key parameters}
\end{center}
\end{table}

\subsection{Sequence}
\label{sec:Sequence}
\begin{figure}[tp]
\centering
\includegraphics*[width=0.95\textwidth]{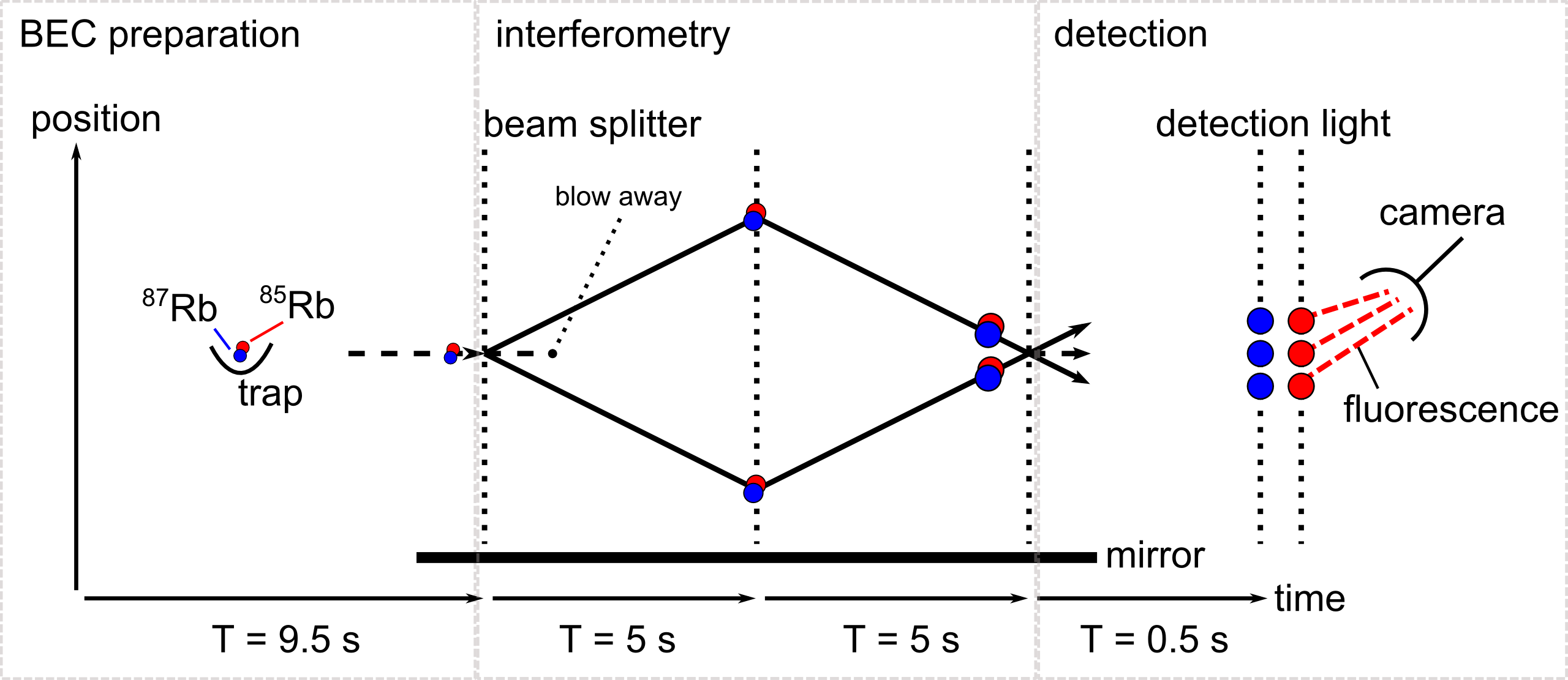}
\caption{Interferometer scheme. The BECs are simultaneously prepared inside the same trap, and coherently manipulated by two synchronous beam splitter light fields. Finally, the output ports of each interferometer are detected at the same time, but subsequently for the two species. The sketch is not to scale.}
\label{fig:int_sequence}
\end{figure}

The measurement sequence breaks down into four steps: BEC generation, state preparation for interferometry, interferometry pulse sequence and detection of the interferometer output ports as depicted in figure~\ref{fig:int_sequence}. The phase information is encoded in the transition probability detected in the end.\\
In the first step, $^{87}$Rb and $^{85}$Rb BECs are simultaneously generated in a miscible regime in the magnetic sensitive states $|F=1, m_{\mathrm{F}}=-1\rangle$ for $^{87}$Rb and $|F=2, m_{\mathrm{F}}=-2\rangle$ for $^{85}$Rb~\cite{Papp2008PRL}. After release from the trap with a final trap isotropic frequency of $42\,\mathrm{Hz}$ the ensembles expand for $80\,\mathrm{ms}$. At this point each cloud contains $10^{6}$ atoms with a velocity spread of $\sigma_{\mathrm{v}}=300\,$\textmu$\mathrm{m/s}$. 
During this expansion time a magnetic offset (Feshbach) field prevents the collapse of the $^{85}$Rb BEC. Subsequently, delta kick cooling~\cite{Dickerson2013PRL,Muntinga2013PRL} is applied by shortly switching the trap on again. This affects both ensembles and further reduces the velocity spread to $\sigma_{\mathrm{v}}=82\,$\textmu$\mathrm{m/s}$. A detailed of the preparation is made in reference~\cite{Posso2013arXiv}.\\
Directly after the kick, the atoms in both ensembles are transferred to  magnetic insensitive states $m_{\mathrm{F}}=0$ via an adiabatic rapid passage. During this procedure a magnetic bias field of $B_{\mathrm{ARP}}=5\,\mathrm{G}$ lifts the degeneracy. This is lowered to $B_{0}=1\,\mathrm{mG}$ afterwards and kept at this value until the end of the whole sequence. At this time  an optional preparation step can be introduced. It is possible to drive the clock transitions $|F=1, m_{\mathrm{F}}=0\rangle\rightarrow|F=2, m_{\mathrm{F}}=0\rangle$ in $^{87}$Rb and $|F=2, m_{\mathrm{F}}=0\rangle\rightarrow|F=3, m_{\mathrm{F}}=0\rangle$ in $^{85}$Rb with a microwave $\pi$-pulse. Afterwards, an optical field driving the transitions $|F=1\rangle\rightarrow|F'=2\rangle$ in $^{87}$Rb and $|F=2\rangle\rightarrow|F'=3\rangle$ in $^{85}$Rb removes atoms not affected by the microwave pulse. These optional two steps allow to enter the interferometer pulse sequence with a different internal state for subsequent cycles. Allowing for some further expansion time of $~1\,\mathrm{s}$, the time needed for the BEC generation and preparation sums up to $9.5\,\mathrm{s}$. The size of the atomic cloud, as given by a Gaussian distribution fit, is $\sigma_{\mathrm{r}}=300\,$\textmu$\mathrm{m}$.\\
Three beam splitter pulses separated by a free evolution time $T=5\,\mathrm{s}$ implement the interferometer pulse sequence using a Raman double diffraction scheme~\cite{Leveque2009PRL,Malossi2010PRA}. First, a $\pi/2$-pulse produces a coherent superposition between the momentum states $p_{0}\pm\hbar k$ originating from the initial momentum state $p_{0}\approx 0$. The internal state also changes, but is the same for $p_{0}\pm\hbar k$. Subsequently, a blow away pulse removes atoms remaining in the initial momentum (and internal) state. After the time $T$, a $\pi$-pulse inverts the momentum. Finally, after $2T$, another $\pi/2$-pulse recombines the trajectories. During the time $2T=10\,\mathrm{s}$ the atoms in the upper and lower interferometer paths propagate in the same internal state. This renders the interferometer immune to AC-Stark shifts and beam splitter phase lock noise. The pulses are applied to both isotopes simultaneously and adjusted to a $\pi/2$-pulse duration of $\tau=50\,$\textmu$\mathrm{s}$ via the intensity of the laser beams. In $^{87}$Rb the states $|F=2, m_{\mathrm{F}}=0\rangle$ and $|F=1, m_{\mathrm{F}}=0\rangle$ and in $^{85}$Rb the states $|F=3, m_{\mathrm{F}}=0\rangle$ and $|F=2, m_{\mathrm{F}}=0\rangle$ are coupled via a Raman transition. To avoid population of the intermediate states, the beam splitter light fields are red detuned by $\Delta_{87}\approx 1.7\,\mathrm{GHz}$ to the $|F'=1, m_{\mathrm{F}}=0\rangle$ state in the case of $^{87}$Rb and by $\Delta_{85}\approx 1.1\,\mathrm{GHz}$ to the $|F'=2, m_{\mathrm{F}}=0\rangle$ state for $^{85}$Rb. This detuning difference is chosen to match the effective wave vectors of both species. After the last pulse for recombination, one output port has the momentum $p_{\mathrm{out,0}}\approx p_{0}$, while the other consists of two momenta $p_{\mathrm{out,}\pm}\approx p_{0}\pm\hbar k$. Due to the preparation, the initial momentum $p_{0}$ relative to the retroreflection mirror is vanishing and the coupling of all
interferometer states can be accomplished with only two retroreflected
laser beams.\\ 
Prior to the detection, a waiting time of $0.5\,\mathrm{s}$ allows the output ports to spatially separate. Then a readout of both output ports of the $^{87}$Rb at the same time is performed. Therefore, cooling light $3\,\mathrm{MHz}$ red detuned to $|F=2\rangle\rightarrow|F'=3\rangle$ in $^{87}$Rb and repumping light $3\,\mathrm{MHz}$ red detuned to $|F=1\rangle\rightarrow|F'=2\rangle$ in $^{87}$Rb are simultaneously applied to the atoms. The signal from the second species $^{85}$Rb is detected subsequently in the same way, addressing the transitions $3\,\mathrm{MHz}$ red detuned to $|F=3\rangle\rightarrow|F'=4\rangle$ in $^{85}$Rb and repumping light $3\,\mathrm{MHz}$ red detuned to $|F=2\rangle\rightarrow|F'=3\rangle$ in $^{85}$Rb. By using this simultaneous read out of both interferometer ports, normalization fluctuations due to detection light frequency and intensity fluctuations are suppressed~\cite{Biedermann2009OL}.
\section{Noise sources, error terms and mitigation}
\label{sec:noise_errors}
The targeted inaccuracy in the E\"otv\"os ratio is $\eta=|\Delta a|/g \leq 2\cdot 10^{-15}$. It will be reached by averaging the shot noise limited signal with a sensitivity $\sigma_{\Delta a}/\sqrt{T_{c}}=(2/N)^{1/2}\cdot(CkT^2)^{-1}=2.93\cdot10^{-12}\,\mathrm{ms}^{-2}$ (see table~\ref{Key parameters}) to differential accelerations $\Delta a=a_{87}-a_{85}$ over a sufficient number of cycles (see section~\ref{sec:performance}). This section discusses the impact and suppression of perturbations onto the differential acceleration signal. 

\subsection{Double diffraction}
As stated in section~\ref{sec:measurement_principle} the interferometer will feature a symmetric beam splitter based on Raman transitions as demonstrated in~\cite{Leveque2009PRL}. In this scheme, both upper and lower path of the interferometer have the same internal state during the free evolution time $2T$, leading to several implications. It inherently suppresses the same phase shifts as the k-reversal technique~\cite{Durfee2006PRL} which requires a steady state to remove systematics. Contrary to this, the suppression for symmetric beam splitters is instantaneous. Consequently, no steady state is required and in addition associated noise figures are also suppressed. The influence of phase noise between the two beam splitting lasers becomes negligible thus relaxing an otherwise very stringent requirement. AC-Stark shifts also vanish. Finally, the symmetric beam splitting does not induce a recoil velocity dependent centre of mass motion and subsequently keeps the centres of mass of the two interferometers superimposed.

\subsection{Mean field}
\label{Mean field}
The inter and intra species interactions can cause a bias in the interferometer signal depending on imperfections of the first beam splitter $\pi/2$ pulse and the density of the BECs~\cite{Debs2011PRA}. For a shot noise limited beam splitting precision with $N$ atoms the time dependent differential frequency between upper and lower interferometer path is
\begin{equation}
 \omega_{\mathrm{mf}}(t)=2\pi\frac{\mu V(0)}{h\sqrt{N}V(t)},
\end{equation}
where $h$ is Planck's constant, $\mu=n(0)U$ the chemical potential with initial peak density $n(0)=N/V(0)$, and $V(t)=4\pi\left(\sigma_{\mathrm{r}}+\sigma_{\mathrm{v}}t\right)^{3}/3$ denotes the volume of the BEC. The interaction parameter $U=ha_{\mathrm{sc}}/(\pi m)$ depends on the scattering length $a_{\mathrm{sc}}$ and atomic mass $m$. Subsequently, the integrated frequency shift $\phi_{\mathrm{mf}}=\int_{0}^{2T}\rmd t\,\omega_{\mathrm{mf}}(t)$ during the free evolution time $2T=10\,\mathrm{s}$ for an initial beam splitting accuracy of $1/\sqrt{N}=0.001$, and division by the scaling factor leads to the acceleration bias $\delta a=\phi_{\mathrm{mf}}/(kT^{2})$. Summing over intra and inter species interactions, the differential acceleration bias $\delta a_{\mathrm{mf}}=\delta a_{87}+\delta a_{85}+\delta a_{87/85}$ can be estimated.\\
The parameters for assessing the bias are: initial volume $V(0)=4\pi(300\,$\textmu$\mathrm{m})^{3}/3$, intra species scattering length $a_{\mathrm{sc,}87}=100\,a_{0}$ for $^{87}$Rb, $a_{\mathrm{sc,}85}=-443\,a_{0}$ for $^{85}$Rb, inter species scattering length $a_{\mathrm{sc,}87/85}=213\,a_{0}$ for $^{87}$Rb/$^{85}$Rb, and the Bohr radius $a_{0}$. Since $a_{\mathrm{sc,}85}$ is negative while the other scattering lengths are positive, the ratio $N_{87}/N_{85}$ between $N_{87}$ $^{87}$Rb and $N_{85}=10^{6}$ $^{85}$Rb atoms can be tuned to inherently minimize $\delta a_{\mathrm{mf}}$. Choosing $N_{87}/N_{85}=1.697$ would lead to an acceleration error of $\delta a_{87}+\delta a_{85}+\delta a_{87/85}=\delta a_{\mathrm{mf}}=-1.7\cdot10^{-16}\,\mathrm{ms}^{-2}$. An uncertainty of $10^{-3}$ in the ratio $N_{87}/N_{85}$ corresponding to $10^{3}$ atoms implies an uncertainty in the bias of $\delta a_{\mathrm{mf}}=2\cdot10^{-15}\,\mathrm{ms}^{-2}$.\\
Fluctuations in total atom number or the ratio will induce a noise contribution. If the ratio is fixed, variations in total atom number of $20\,\%$ lead to negligible fluctuations in the bias of $4\cdot10^{-16}\,\mathrm{ms}^{-2}$. More important, fluctuations of the ratio $N_{87}/N_{85}$ by $20\,\%$ for fixed atom number $N_{85}$ lead to a noise contribution of $3\cdot10^{-13}\,\mathrm{ms}^{-2}$ per cycle, roughly a factor of $10$ below the shot noise limit.

\subsection{Linear accelerations}
\label{Linear accelerations}
Due to the scaling factor $kT^{2}\approx 8\cdot10^{8}\,\mathrm{s}^2\mathrm{m}^{-1}$ random vibrations parallel to the effective wave vectors will wash out the fringe visibility in a single interferometer. If the correlation between the two interferometers is sufficiently high it is still possible to extract the differential acceleration with ellipse fitting methods~\cite{Foster2002OL,Stockton2007PRA}. This implies a matched scaling factor $k_{87}T_{87}^{2}=k_{85}T_{85}^{2}$. Maximum correlation is reached if the effective wave vectors are the same $k_{87}\approx k_{85}$, both atomic ensembles interact with the beam splitting light fields at the same time, and the effective Rabi frequencies are matched $\Omega_{87}=\Omega_{85}$. The latter can be understood by considering the transfer function $H(f)$~\cite{Cheinet2008IEEE} which translates the power spectral density $S(f)$ of random zero-mean accelerations into rms phase noise of the interferometer
\begin{equation}
 \sigma_{\phi}^{2}=\frac{1}{2}\frac{1}{m^{2}}\int_{0}^{\infty}\rmd f\,4\frac{\sin^{4}(2\pi mT_{\mathrm{c}}/2)}{\sin^{2}(2\pi T_{\mathrm{c}}/2)}|H(f)|^2S(f).
\end{equation}
This equation assumes an integration over $m$ interferometer cycles with a duration $T_{\mathrm{c}}$. In the differential interferometer, the transfer function is rewritten as
\begin{equation}
 |H_{\Delta \mathrm{a}}(f)|^2=|H_{\mathrm{a}87}(f)-H_{\mathrm{a}85}(f)|^2
\end{equation}
with
\begin{eqnarray*}
 H_{\mathrm{a}}(f) & = & \frac{k}{(2\pi f)^2}\frac{4i\Omega_{\mathrm{R}}}{(2\pi f)^2-\Omega_{\mathrm{R}}^2}\sin\left( \frac{2\pi f(T+2\tau)}{2} \right) \\
 & \cdot & \left( \cos\left( \frac{2\pi f(T+2\tau)}{2} \right)+\frac{\Omega_{\mathrm{R}}}{2\pi f}\sin\left( \pi fT \right) \right).
\end{eqnarray*}
Here, $\tau$ denotes the duration of a $\pi /2$ beam splitter pulse.\\
Matching the wave vectors to $\Delta k/k\approx10^{-9}$, the Rabi frequencies to $\Delta \Omega_{\mathrm{R}}/\Omega_{\mathrm{R}}\approx10^{-4}$, and assuming $\tau_{87}=\tau_{85}\approx50\,$\textmu$\mathrm{s}\approx\pi/(4\Omega_{\mathrm{R}})$ leads to a suppression ratio of $2.5\cdot10^{-9}$. Thus, spurious bias accelerations of the apparatus of up to $4\cdot10^{-7}\,\mathrm{ms}^{-2}$ would be suppressed in the differential signal below $10^{-15}\,\mathrm{ms}^{-2}$. A power spectral density of acceleration noise of up to $5\cdot10^{-4}\,\mathrm{ms}^{-2}\mathrm{Hz}^{-1/2}$ would lead to a rms noise in the differential signal of $\approx10^{-12}\,\mathrm{ms}^{-2}$ compatible with the shot noise limited sensitivity for the instrument.\\
An additional constraint is to keep the Doppler shift smaller than the Rabi frequency which translates into $k\cdot a_{\mathrm{rms}}/(2\pi f)<\Omega_{\mathrm{R}}\approx1.6\,\mathrm{kHz}$. Thus, the interferometer works with a power spectral density of acceleration noise of up to to $10^{-3}f\,\mathrm{Hz}^{-1}\mathrm{ms}^{-2}\mathrm{Hz}^{-1/2}$ for frequencies $f$ in Hz.

\subsection{Rotations and gravity gradients}
\label{Rotations and gravity gradients}
Rotations and gravity gradients can accelerate the atomic ensembles with respect to the reference mirror and consequently cause bias terms. These however have the same suppression ratio as discussed in section~\ref{Linear accelerations}. Remaining terms mostly depend on the overlap of the two atomic ensembles. They are derived following~\cite{Borde2004GRG} and~\cite{Hogan2009proc}. The total phase shift corresponds to the sum over the three atom light interactions which in principle compare the center of mass position of the atomic trajectories to the reference mirror. With the position of the upper [lower] path $\vec{q}_{\mathrm{un}}$ [$\vec{q}_{\mathrm{ln}}$] and the effective wave vector acting on the upper [lower] path $\vec{k}_{\mathrm{un}}$ [$\vec{k}_{\mathrm{ln}}$] at the n-th pulse the equation takes the form~\cite{Borde2004GRG}
\begin{equation}
 \Phi=\sum_{n=1}^{3}\left[ \left( \vec{k}_{\mathrm{un}}-\vec{k}_{\mathrm{ln}} \right) \frac{\vec{q}_{\mathrm{ln}}+\vec{q}_{\mathrm{un}}}{2} \right]
\end{equation}
for $|\vec{k}_{\mathrm{un}}|=|\vec{k}_{\mathrm{ln}}|=k/2$. This equation includes all three contributions by the laser interaction, the action integral along each path, and the separation at the last beam splitter pulse. By solving the equations of motion with a polynomial ansatz the positions are calculated following~\cite{Hogan2009proc}. Resulting acceleration bias terms $\Delta a=\phi/(kT^{2})$ are reported in table~\ref{tab:rot_grad_bias}.
\begin{table}[tb]
\begin{center}
 \caption{Differential acceleration biases depending on initial overlap and velocity difference. Other terms are at least one order of magnitude smaller than the target and thus negligible. The evaluation adopted a perigee at $700\,\mathrm{km}$ where $\Omega_{\mathrm{orb}}=1.4\,\mathrm{rad/s}$, $T_{\mathrm{zz}}=2.2\cdot10^{-6}\,\mathrm{s}^{-2}$, and $T_{\mathrm{xx}}=T_{\mathrm{yy}}=-T_{\mathrm{zz}}/2$. The inertial pointing mode is implemented via the counter rotation $\Omega_{\mathrm{c}}\approx\Omega_{\mathrm{orb}}$ with a mismatch of $1\,$\textmu$\mathrm{rad/s}$. Residual rotations are $\Omega_{\mathrm{x}}=\Omega_{\mathrm{y}}=\Omega_{\mathrm{z}}=1\,$\textmu$\mathrm{rad/s}$.}
 \begin{tabular}{llc}\hline
   Phase shift $\phi$ & Differential velocity $\Delta v_{\mathrm{r}}$ & Differential \\
  & / spatial separation $\Delta r$ & acceleration \\
  & & in $10^{-15}\,\mathrm{ms}^{-2}$ \\ \hline
  $kT^{3}T_{\mathrm{xx}}\Omega_{\mathrm{orb}}\Delta x$ & $\Delta x=1.1\,\mathrm{nm}$ & $9.1\cdot10^{-3}$ \\ 
  $kT^{3}(2\Omega_{\mathrm{orb}}^{3}+\Omega_{\mathrm{c}}^{3})\Delta x$ & $\Delta x=1.1\,\mathrm{nm}$ & $4.9\cdot10^{-2}$ \\ 
  $-2kT^{2}\Omega_{\mathrm{y}}\Delta v_{\mathrm{x}}$ & $\Delta v_{\mathrm{x}}=0.31\,\mathrm{nm/s}$ & $-6.3\cdot10^{-1}$ \\ 
  $-7/6\cdot kT^{4}T_{\mathrm{zz}}\Omega_{\mathrm{orb}}\Delta v_{\mathrm{x}}$ & $\Delta v_{\mathrm{x}}=0.31\,\mathrm{nm/s}$ & $2.9\cdot10^{-2}$ \\ 
  $-7/6\cdot kT^{4}T_{\mathrm{xx}}\Omega_{\mathrm{orb}}\Delta v_{\mathrm{x}}$ & $\Delta v_{\mathrm{x}}=0.31\,\mathrm{nm/s}$ & $-1.5\cdot10^{-2}$ \\ 
  $-kT^{2}\Omega_{\mathrm{orb}}\Omega_{\mathrm{z}}\Delta y$ & $\Delta y=1.1\,\mathrm{nm}$ & $-6.3\cdot10^{-1}$ \\ 
  $-2kT^{2}\Omega_{\mathrm{x}}\Delta v_{\mathrm{y}}$ & $\Delta v_{\mathrm{y}}=0.31\,\mathrm{nm/s}$ & $-6.3\cdot10^{-1}$ \\ 
  $-kT^{2}T_{\mathrm{zz}}\Delta z$ & $\Delta z=1.1\,\mathrm{nm}$ & $2.6$ \\ 
  $-kT^{2}(\Omega_{\mathrm{orb}}^{2}-\Omega_{\mathrm{c}}^{2})\Delta z$ & $\Delta z=1.1\,\mathrm{nm}$ & $3.2\cdot10^{-3}$ \\ 
  $-kT^{3}T_{\mathrm{zz}}\Delta v_{\mathrm{z}}$ & $\Delta v_{\mathrm{z}}=0.31\,\mathrm{nm/s}$ & $3.5$ \\ 
  $-3kT^{3}(2\Omega_{\mathrm{c}}\Omega_{\mathrm{orb}}-\Omega_{\mathrm{orb}}^{2}-\Omega_{\mathrm{c}}^{2})\Delta v_{\mathrm{z}}$ & $\Delta v_{\mathrm{z}}=0.31\,\mathrm{nm}$ & $<10^{-3}$ \\ \hline
 \end{tabular}
 \label{tab:rot_grad_bias}
 \end{center}
 \end{table}
The different recoil velocities and masses of $^{87}$Rb and $^{85}$Rb give rise to a bias in differential acceleration dependent on the second order gravity gradient $T_{\mathrm{zzz}}$
\begin{equation}
 \Delta a=\left( -\frac{T^{4}T_{\mathrm{zzz}}\hbar^{2}k^{3}}{16m_{87}^{2}}-\left(-\frac{T^{4}T_{\mathrm{zzz}}\hbar^{2}k^{3}}{16m_{85}^{2}} \right) \right)\cdot\frac{1}{kT^{2}}. \nonumber
\end{equation}
Considering Earth's contribution $T_{\mathrm{zzz}}=9.6\cdot10^{-13}\,\mathrm{m}^{-1}\mathrm{s}^{-2}$ at an altitude of $700\,\mathrm{km}$ this term yields $3.9\cdot10^{-17}\,\mathrm{ms}^{-2}$. Phase terms $\propto \hbar k^{2}/m$ are suppressed due to symmetry of the double diffraction scheme.\\
Self-gravity arising from the payload itself~\cite{DAgostino2011Met} strongly depends on the instrument design and should inherently be reduced by a choosing a symmetric mass distribution of and around the vacuum chamber. Second order gradients can still imply a considerable bias which can be suppressed by alternating the sensitive axis (see section~\ref{sec:orbit}) since they should be stable in time.

\subsection{Magnetic fields}
\label{sec:Magnetic_fields}
Magnetic fields are used to capture and cool the atomic ensembles, to tune repulsive interactions between atoms via Feshbach resonances, and lift the degeneracy~\cite{Posso2013arXiv}. Gradients of the magnetic fields during the preparation will affect the overlap which subsequently can cause a bias (see section~\ref{Rotations and gravity gradients}). A magnetic field gradient during the interferometer pulse sequence will directly cause a differential bias.  
\paragraph{Preparation for interferometry} The critical part of the preparation is the time after release from the trap. At this point, the atoms are in magnetic sensitive sub states $|F=1,m_{\mathrm{F}}=-1\rangle$ for $^{87}$Rb, $|F=2,m_{\mathrm{F}}=-2\rangle$ for $^{85}$Rb and the Feshbach field $B_{\mathrm{F}}=155\,\mathrm{G}$ is switched on. A magnetic field gradient $\delta B$ couples to the linear and quadratic Zeeman effect and exhibits a force~\cite{Steck2008_Rb87,Steck2008_Rb85}. This force is different for the two isotopes due to different internal states and atomic properties. The linear Zeeman effect leads to a non vanishing relative acceleration
\begin{equation}
 \Delta b_{\mathrm{lin}} = b_{85\mathrm{,lin}}-b_{87\mathrm{,lin}} = m_{\mathrm{F,}85}\mu_{\mathrm{B}}g_{\mathrm{f,}85}\frac{\delta B}{m_{85}}-m_{\mathrm{F,}87}\mu_{\mathrm{B}}g_{\mathrm{f,}87}\frac{\delta B}{m_{87}}, \nonumber
\end{equation}
where $m_{\mathrm{F}}$ denotes the magnetic sub state, $\mu_{\mathrm{B}}$ the Bohr magneton, $g_{\mathrm{f}}$ the Land{\'e} factor, and $m$ the atomic mass. Additionally, the differential acceleration due to the quadratic Zeeman effect can be calculated according to~
\begin{equation}
 \Delta b_{\mathrm{q}} = b_{85\mathrm{,q}}-b_{87\mathrm{,q}} = hK_{85}B_{\mathrm{F}}\frac{\delta B}{m_{85}}-hK_{87}B_{\mathrm{F}}\frac{\delta B}{m_{87}},
 \label{eqn:quad_Zeeman}
\end{equation}
with Planck's constant $h$, clock transition Zeeman shift $K_{87}=575.15\,\mathrm{Hz/G}^2$ for $^{87}$Rb and $K_{85}=1293,98\,\mathrm{Hz/G}^2$ for $^{85}$Rb. From these differential accelerations the differential position and velocity at the first beam splitter pulse can be derived. Assuming a time $t_{\mathrm{F}}=0.1\,\mathrm{s}$ between release and transfer to the non magnetic states / Feshbach field switch off, the requirement on the magnetic field gradients in all three axes is $\delta B < 4\cdot10^{-6}\mathrm{G/m}$ (see Table~\ref{tab:preparation_magnetic_field_gradients}). Propagation of the two ensembles in $m_{\mathrm{F}}=0$ states for $1\,\mathrm{s}$ in an offset field of $B_{0}=1\,\mathrm{mG}$ leads to negligible contributions of $\Delta v_{0}=0.1\,\mathrm{pm/s}$ and $\Delta r_{0}=0.01\,\mathrm{pm}$ even for a gradient of $\delta B=10^{-4}\,\mathrm{G/m}$.
\begin{table}[t]
\begin{center}
\caption{Requirements on magnetic field gradients after release from the ODT until transfer into non magnetic internal states and Feshbach field switch off. The requirements are driven by the requirements on differential velocity and position at the first beam splitter pulse (see Table~\ref{tab:rot_grad_bias}).}
\begin{tabular}[top]{llc}
\hline
Quantity & Limit & $\delta B$ in G/m \\
  &  & all three axes \\ \hline
Differential velocity & $\Delta v<0.31\,\mathrm{nm/s}$ & $4\cdot10^{-6}$  \\
Differential position & $\Delta r<1.1\,\mathrm{nm}$ & $3.2\cdot10^{-4}$  \\ \hline 
\end{tabular}
\label{tab:preparation_magnetic_field_gradients}
\end{center}
\end{table}
\paragraph{Interferometer pulse sequence} Before starting the interferometric sequence, the atoms are transferred to the $m_{\mathrm{F}}=0$ sub states and a magnetic offset field $B_{\mathrm{O}}=1\,\mathrm{mG}$ lifts the degeneracy. If a magnetic field gradient is present a differential acceleration results according to equation~\ref{eqn:quad_Zeeman}. To keep the differential bias $\Delta a_{\mathrm{B}_{\mathrm{O}}}<10^{-15}\,\mathrm{ms}^{-2}$ the requirement on the magnetic field gradient during the beam splitter interval and along the sensitive direction (z-axis) is $\delta B<1\,$\textmu$\mathrm{G/m}$. A mitigation strategy is to alternate the hyperfine levels of both isotopes at the interferometer input for subsequent cycles. This changes the sign of the bias acceleration $\Delta a_{\mathrm{B}_{\mathrm{O}}}$ and summing over two subsequent interferometer signals is expected to suppress the bias by a factor of $\sim 500$.

\subsection{Differential displacement inside the trap}
Prior to the release from the trap, gravity gradients, rotations, magnetic field gradients, and bias accelerations can induce a differential displacement $\Delta r$ of the two isotopes inside the trap~\cite{Hogan2009proc}. This $\Delta r$ has to be below the requirements stated in table~\ref{tab:preparation_magnetic_field_gradients}.\\
If the trap is displaced to the center of mass of the spacecraft by a distance $r$ the force imposed by a gravity gradient $T_{\mathrm{rr}}$ leads to a displacement $d$ until equilibrium with the restoring force of the trap with frequency $\omega$ is reached:
\begin{eqnarray}
2m_{85}T_{\mathrm{rr}}r & = & m_{85}d_{85}\omega_{85}^{2} \\ \nonumber
2m_{87}T_{\mathrm{rr}}\left(r+\Delta r\right) & = & m_{87}\left(d_{87}+\Delta r\right)\omega_{87}^{2} \\ \nonumber
\Rightarrow \Delta r & = & \frac{2T_{\mathrm{rr}}r\left( m_{85}/m_{87} -1\right)}{2T_{\mathrm{rr}}-\omega_{87}^{2}}.
\label{eqn:grav_sag}
\end{eqnarray}
Given a trap frequency of $\omega = 42\,\mathrm{Hz}$ as proposed in~\cite{Posso2013arXiv}, a maximum gravity gradient $T_{\mathrm{rr}}=2.2\cdot10^{-6}\,\mathrm{s}^{-2}$ (table~\ref{tab:orbit_parameters}), and a distance to the center of mass $r < 2\,\mathrm{m}$, the resulting differential displacement is  in $\Delta r = 120\,\mathrm{pm}$.\\
Rotations $\Omega$ of the experiment platform lead to the same effect. For the estimation $2T_{\mathrm{rr}}$ in equation~\ref{eqn:grav_sag} is substituted by $\Omega^{2}$. With the maximum rotation rate $\Omega=1.4\,\mathrm{mrad/s}$ from table~\ref{tab:orbit_parameters} the differential displacement is $\Delta r = 53\,\mathrm{pm}$.\\
As described in section~\ref{sec:Magnetic_fields} magnetic field gradients will exert a different force onto the two ensembles, implying a differential displacement
\begin{equation}
\Delta r=\frac{hK_{87}B_{\mathrm{F}}+\mu_{\mathrm{B}}}{m_{87}\omega}\delta B -\frac{hK_{85}B_{\mathrm{F}}+\mu_{\mathrm{B}}}{m_{85}\frac{m_{85}}{m_{87}}\omega}\delta B. \nonumber
\end{equation}
To keep the differential displacement below $\Delta r \leq 250\,\mathrm{pm}$ a gradient $\delta B \leq 12\,$\textmu$\mathrm{G/m}$ suffices which is less a restrictive requirement with respect to that stated in section~\ref{sec:Magnetic_fields}.\\
Another possible source for a differential displacement inside the trap is the bias acceleration $a_{\mathrm{bias}}$ of the experiment platform
\begin{equation}
\Delta r=\frac{a_{\mathrm{bias}}}{\omega ^2}\frac{m_{87}-m_{85}}{m_{87}}. \nonumber
\end{equation}
A differential estimated achievable displacement $\Delta r= 250\,\mathrm{pm}$ can be estimated for $a_{\mathrm{bias}}=20\,$\textmu$\mathrm{ms}^{-2}$ which is above the drag stated in table~\ref{tab:orbit_parameters} and thus compatible.\\
Summing up, the displacement is $\Delta r\approx 0.7\,\mathrm{nm}$, below the requirement of $1.1\,\mathrm{nm}$ stated in table~\ref{tab:rot_grad_bias}.

\subsection{Wave fronts}
The effective wavefront of the beam splitter lasers is the reference for the position of the atoms at each of the three atom-light interactions. It is defined by the retroreflection mirror which reflects one of the two light fields forming the beam splitter for each species, and the initial collimation when both light fields are still superimposed.\\
To estimate the impact of non ideal retroreflection optics a defocus leading to an effective wave front curvature $R$ is evaluated~\cite{Louchet2011NJP}. Coupled to the atomic temperatures $T_{\mathrm{at}}$ a differential acceleration bias arises
\begin{equation}
 \Delta a_{\mathrm{wf,r}}=\frac{k_{\mathrm{B}}}{R}\left(\frac{T_{\mathrm{at,}87}}{m_{\mathrm{at,}87}}-\frac{T_{\mathrm{at,}85}}{m_{\mathrm{at,}85}}\right), \nonumber
\end{equation}
where $k_{\mathrm{B}}$ denotes Boltzmann's constant and $m$ the atomic mass. Assuming an effective atomic temperature $T_{\mathrm{at,}87}=T_{\mathrm{at,}85}\approx 70\,\mathrm{pK}$ and $R=250\,\mathrm{km}$ corresponding to an object with a surface flatness of $\lambda/300$ implies $\Delta a_{\mathrm{rro}}=6.3\cdot10^{-16}\,\mathrm{ms}^{-2}$.\\
All beam splitter light fields are guided by the same optical single mode fibre to the experiment and collimated by the same telescope. The retroreflection setup then suppresses wave front errors from the telescope optics. Still, the finite initial collimation quality with wave front curvature $R_{\mathrm{i}}$ results in an effective wave front curvature which leads to a differential acceleration bias
\begin{eqnarray}
\label{eqn:wf_initial_collimation}
 \Delta a_{\mathrm{wf,c}} & = & \frac{\phi_{\mathrm{wf,c}}}{kT^{2}} = \frac{1}{kT^{2}}\cdot k\left[ \left(\sigma_{\mathrm{r}} +\sigma_{\mathrm{v}} t_{0} \right)^{2} \left( \frac{1}{R_{\mathrm{i,c}}}-\frac{1}{R_{\mathrm{b,c}}} \right) \right. \\ \nonumber
 & - & \left( \left(\sigma_{\mathrm{r}} +\sigma_{\mathrm{v}} \left(t_{0}+T\right) \right)^{2} \left( \frac{1}{R_{\mathrm{i,t}}}-\frac{1}{R_{\mathrm{b,c}}} \right) \right. \\ \nonumber
 & + & \left. \left(\sigma_{\mathrm{r}} +\sigma_{\mathrm{v}} \left(t_{0}+T\right) \right)^{2} \left( \frac{1}{R_{\mathrm{i,m}}}-\frac{1}{R_{\mathrm{b,m}}} \right) \right) \\ \nonumber
 & + & \left. \left(\sigma_{\mathrm{r}} +\sigma_{\mathrm{v}} \left(t_{0}+2T\right) \right)^{2} \left( \frac{1}{R_{\mathrm{i,c}}}-\frac{1}{R_{\mathrm{b,c}}} \right) \right].
\end{eqnarray}
This calculation takes into account the position dependence of the initial wave front curvature $R_{\mathrm{i,}n}$. Here, $n=c$ denotes the initial (center) position at the first beam splitter pulse, $n=t$ the position near the telescope, and $n=m$ the position near the retro reflection mirror. These latter two positions will slightly differ for $^{87}$Rb and $^{85}$Rb because of their different recoil velocities for a single photon transition $v_{\mathrm{r}87}=5.8845\,\mathrm{mm/s}$ and $v_{\mathrm{r}85}=6.023\,\mathrm{mm/s}$~\cite{Steck2008_Rb85,Steck2008_Rb87}. Starting with a numerical aperture of the fibre guiding the light fields to the telescope $NA = 0.12$ and a collimation lens with focal length $f = 0.2\,\mathrm{m}$, the wave front curvature and beam waist are calculated at the relevant distance from the collimation lens. Alignment with a precision of $100\,$\textmu$\mathrm{m}$ implying a distance of $19.99\,\mathrm{cm}$ between fibre and lens leads to a wave front curvature of $R_{\mathrm{i}}\approx 400\,\mathrm{m}$ and a waist of $19.7\,\mathrm{mm}$. This corresponds to effective (differential) wave front curvatures above $900\,\mathrm{km}$. Telescopes creating wave front curvatures of $400\,\mathrm{m}$ are routinely operated in lab based experiments~\cite{Tackmann2012NJP}. Evaluation of equation~\ref{eqn:wf_initial_collimation} estimates the differential acceleration bias to $\Delta a_{\mathrm{wf,c}}=2.8\cdot10^{-16}\,\mathrm{ms}^{-2}$. \\
The requirements onto the effective wave front curvature can be relaxed by matching the expansion rates of $^{87}$Rb and $^{85}$Rb. According to simulations in~\cite{Posso2013arXiv} a match of $\approx10^{-3}$ could be possible. This would reduce the requirements by a factor of $\approx20$.

\subsection{Beam splitter laser frequency stability}
Beam splitter laser frequency jitter affects the noise background of the interferometer. Critical parts are the detection lasers and beam splitter lasers. During the beam splitting process the two laser light fields driving the Raman transition travel the same optical path except for the retro reflection. This causes a delay line for one of the beams according to twice the distance atoms - retro reflection mirror. The impact of this effect is described in~\cite{LeGouet2007EPJD}. To estimate the noise contribution a constant delay line $s_{\mathrm{d}}=30\,\mathrm{cm}$ implying a delay time $t_{\mathrm{d}}=s_{\mathrm{d}}/c=1\,\mathrm{ns}$ with speed of light $c$, a $\pi/2$-pulse duration of $\tau=50\,$\textmu$\mathrm{s}$, and a spectral density for white frequency noise $S_{\nu}^{0}=32\cdot10^{4}\,\mathrm{Hz}^{2}\mathrm{/Hz}$ corresponding to a Lorentzian linewidth of $100\,\mathrm{kHz}$~\cite{Riehle2004FS} are assumed. This leads to a noise contribution by a factor of $4$ below the shot noise limit.

\subsection{Detection} As described in section~\ref{sec:Sequence} the interferometer output ports of a single species will be read out at the same time by simultaneous illumination with the same laser light field and fluorescence signal detection. Frequency jitter of the illumination laser is common mode in both output ports and drops out after normalization. Using a comparable approach, close to shot noise limited detection for $~10^{8}$ cesium atoms was demonstrated in~\cite{Biedermann2009OL}.\\
As part of the error budget the detection efficiencies of the two output ports of a single interferometer have to be considered. The normalized output signal can be written as $P=S_{1}/(S_{1}+\epsilon S_{2})$, where $S_{1}$ and $S_{2}$ denote the atomic signals, and $\eta$ is the differential readout efficiency between the ports $1$ ($p_{\mathrm{out,}\pm}\approx p_{0}\pm\hbar k$) and $2$ ($p_{\mathrm{out,}0}\approx p_{0}$). In~\cite{Sorrentino13arXiv} the error in an ellipse fit for $\eta\neq 1$ was simulated and a quadratic dependence of interferometer phase $\phi$ on $\epsilon-1$ for $|\epsilon-1|\ll 1$ observed. Extrapolation from the stated results imposes the requirement of $|\epsilon-1|<0.003$ to keep the error in differential acceleration below $10^{-15}\,\mathrm{ms}^{-2}$.

\section{Interferometer contrast}
\label{sec:Interferometer_contrast}
The measurement of the interferometer phase and thus the acceleration is performed indirectly by determining the transition probability at the end of the interferometer sequence. This explains the importance of a high interferometer contrast, since its reduction determines a linear loss of the measurement's sensitivity. In this paragraph the dominant contributions to a contrast loss are described.\\
Contributions due to single photon transitions are negligible due to the detuning of above $1\,\mathrm{GHz}$. Contrast reduction due to velocity selectivity of the beam splitter is expected to be negligible. Due to the Fourier width of the beam splitter corresponding to a temperature of $370\,\mathrm{nK}$~\cite{Steck2008_Rb87,Steck2008_Rb85} which is large against the clouds effective temperature of $70\,\mathrm{pK}$, we can expect a nearly perfect beam splitter efficiency and thus no contrast reduction.\\
Assuming a simple detection scheme, the transition probability is read out by averaging over the whole cloud. Then, a spatial dependent phase $\phi(r_{\mathrm{s}},v_{\mathrm{s}})$ in the detected ensemble leads to contrast reduction due to the spatial $f(r_{\mathrm{s}})$ and velocity spread $g(v_{\mathrm{s}})$ of the atomic ensemble with a resulting spatial spread for the transition probability~\cite{Tackmann2012NJP}. The effective signal with amplitude $A$, amplitude offset $P_{0}$ and phase $\phi(r,v)$ at mean position $r$ and with mean velocity $v$ can be calculated by integrating the expected signal $P_{\mathrm{s}}$ for the position and velocity dependent phases $\phi(r_{\mathrm{s}},v_{\mathrm{s}})$:
\begin{equation}
P_{\mathrm{tot}}=\int_{-\infty}^{\infty}\rmd r_{\mathrm{s}}\,\rmd v_{\mathrm{s}}\, f(r_{\mathrm{s}})g(v_{\mathrm{s}}) P_{s} (\phi(r_{\mathrm{s}},v_{\mathrm{s}}))=P_{0}-A(\sigma_{\mathrm{r}},\sigma_{\mathrm{v}})\cos{\phi(r,v)}, 
\end{equation}
with $f(r_{\mathrm{s}})$ and $g(v_{\mathrm{s}})$ as the respective distribution functions. The distributions are assumed as Gaussian and the expansion of the atomic ensemble as linear with standard deviation $\sigma_{\mathrm{r}} = 300\,$\textmu$\mathrm{m}$ for $f(r_{\mathrm{s}})$ and $\sigma_{\mathrm{v}} = (k_{\mathrm{B}}T_{\mathrm{at}}/m_{\mathrm{at}})^{-1/2}=82\,$\textmu$\mathrm{m/s}$ for $g(v_{\mathrm{s}})$ with Boltzmann's constant $k_{\mathrm{B}}$, atomic temperature $T_{\mathrm{at}}$ and atomic mass $m_{\mathrm{at}}$. The atomic velocity couples to the rotation of the experimental platform $\Omega$ and leads to a velocity dependent phase shift via the Sagnac effect. For the estimated residual rotation $\Omega=1\,$\textmu$\mathrm{rad/s}$ a contrast $C=A/P_{0}\approx 2A>0.96$ can be expected. At the same time the gravity gradient due to Earth leads to a position dependent acceleration and thus phase shift for different atoms in the cloud. The main contribution arises along the beam splitter axis. Close to Earth at perigee pass with an altitude of $700\,\mathrm{km}$, the gravity gradient $T_{\mathrm{zz}}(700\,\mathrm{km})=2.2\cdot10^{-6}\,\mathrm{s}^{-2}$ reduces the contrast to $C=0.6$. Other gravity gradient components can only couple through rotations and are negligible for the proposed experimental setup. The same holds true for the projection of the other gravity gradients to the $T_{\mathrm{zz}}$ component and will be disregarded. Based on this the main contribution for contrast reduction is the gravity gradient $T_{\mathrm{zz}}$.\\
Recent publications show the possibility to mitigate the loss of contrast induced by a phase gradient over the cloud. By resolving the spatial structure of the interferometer ports and fitting their profiles, this point source interferometry technique retains contrast that would be lost by just integrating over the output ports~\cite{Dickerson2013PRL,Muntinga2013PRL}. An implementation of such a fitting routine seems possible in a space borne experiment, but is not the baseline in this discussion.

\section{Performance versus orbit choice}
\label{sec:performance}
The orbit choice affects the sensitivity onto the E\"otv\"os parameter since the local gravitational acceleration $\vec{g}$ is dependent on the altitude. Additionally, the projection of the gravity gradient onto the sensitive axis also changes which has implications for the interferometer contrast. This leads to an integration behaviour for $N$ measurements described by:
\begin{equation}
 \sigma^{\mathrm{int}}_{\eta}=\sqrt{\frac{1}{N-1} \sum^{N}_{n=1}\left( \frac{\sqrt{2}}{\sqrt{N_{\mathrm{at}}}}\frac{1}{C\left( r_{\mathrm{orb}}(t),\Theta(t) \right) }\frac{1}{\vec{k}\cdot\vec{g}\left( r_{\mathrm{orb}}(t),\Theta(t) \right)T^2}\right)^{2} }.
\end{equation}
Herein, the atom numbers $N_{\mathrm{at}}=N_{87}=N_{85}$ and contrast $C$ are the same for both isotopes. Contrast and projection of local gravitational acceleration $\vec{g}$ onto the sensitive axis defined by the wave vector $\vec{k}$ depend on the distance to Earth $r_{\mathrm{orb}}(t)$ and the attitude towards Earth $\Theta(t)$ which itself is time dependent. At the minimum perigee altitude of $700\,\mathrm{km}$, a sensitivity to the E\"otv\"os ratio of $\sigma^{700\,\mathrm{km}, N=75}_{\eta}=5\cdot10^{-14}$ is reached per revolution after averaging over $75$ interferometer cycles around perigee. Performing more measurements per revolution does not increase the sensitivity since the projection of local gravitational acceleration becomes too small. For a maximum perigee altitude of $2200\,\mathrm{km}$, a similar sensitivity to the E\"otv\"os ratio of $\sigma^{2200\,\mathrm{km}, N=100}_{\eta}=5.3\cdot10^{-14}$ is reached per revolution after averaging over $100$ interferometer cycles. Indeed, the reduced value of local gravitational acceleration is partly compensated by the higher contrast available at higher altitudes. Taking into account the drifting orbit, the integrated signal leads to sensitivity of $\sigma^{1.5\,\mathrm{y}}_{\eta}=2\cdot10^{-15}$ after $1.5\,\mathrm{years}$, in line with the mission life time of $5\,\mathrm{years}$.
\section{Error budget}
The results from section~\ref{sec:noise_errors} are reported in table~\ref{tab:statistical_errors_table} and table~\ref{tab:error_budget_table}. Statistical errors are treated as uncorrelated and are compatible with a shot noise limited sensitivity to differential accelerations of $3.2\cdot10^{-12}\,\mathrm{ms}^{-2}$. This is enabled by the suppression ratio for spurious accelerations and vibrations which contribute a similar noise term $\approx10^{-12}\,\mathrm{ms}^{-2}$ as the beam splitter laser linewidth. Other noise terms are at least by a factor of $3$ below shot noise.\\
 \begin{table}[tb]
 \begin{center}
 \caption{Assessment of statistical errors for the dual species atom interferometer with $^{87}$Rb and $^{85}$Rb per cycle of $20\,\mathrm{s}$. All contributions are expected to be uncorrelated. The dominant noise source is the shot noise.}
 \begin{tabular}{p{0.3\textwidth}p{0.45\textwidth}c}\hline
   \textbf{Noise source} & \textbf{Conditions} & \textbf{Limit in $\mathrm{ms}^{-2}$} \\ \hline
  Shot noise & $N=10^{6}$, $C=0.6$ & $2.93\cdot10^{-12}$ \\ 
  Linear vibrations & Suppression ratio $2.5\cdot10^{-9}$ & $\approx10^{-12}$  \\ 
  Beam splitter & Linewidth $100\,\mathrm{kHz}$  & $8\cdot10^{-13}$  \\ 
  Magnetic fields & $B_{0}=1\,\mathrm{mG}$, $\nabla B_{0}=83\,$\textmu$\mathrm{G/m}$ & $1.1\cdot10^{13}$  \\ 
  Mean field & Beam splitting accuracy $0.001$, & $3\cdot10^{-13}$  \\ 
   & $20\,\%$ fluctuation in $N_{87}/N_{85}$ &  \\
  Overlap & $10\,\%$ fluctuation per cycle & $<10^{-13}$  \\ \hline 
  Sum &  & $3.2\cdot10^{-12}$  \\ \hline
 \end{tabular}
 \label{tab:statistical_errors_table}
 \end{center}
 \end{table}
Uncertainties in the bias terms add up to $7.9\cdot10^{-15}\,\mathrm{ms}^{-2}$ at perigee with an altitude of $700\,\mathrm{km}$. Herein, most uncertainties are expected to be uncorrelated except for those dependent on the overlap and the effective wave front curvature. Consequently, contributors in these subsets are summed up linearly. The uncertainty in the E\"otv\"os ratio of $1\cdot10^{-15}$ is obtained by dividing the bias uncertainty by the projection of local gravitational acceleration onto the sensitive axis of $8\,\mathrm{ms}^{-2}$. It is dominated by the uncertainty of the overlap coupled to gravity gradients and spurious rotations. For higher altitudes and different attitudes, the projections of local gravitational acceleration and Earth's gravity gradient onto the sensitive axis both change. These two effects partly compensate each other: At perigee with an altitude of $2200\,\mathrm{km}$ the uncertainty in the E\"otv\"os ratio remains at $1\cdot10^{-15}$ while it increases to $2\cdot10^{-15}$ at the edges of the perigee arc, still within the targeted inaccuracy. The necessary overlap quality will be verified by spatially resolved imaging after different times of flight. \\
Except for spurious accelerations of the spacecraft which is mitigated by the suppression ratio, remaining terms are related to the payload itself. Requirements on the magnetic field gradient during interferometry and the curvature of the effective beam splitter wave front are relaxed by alternating the internal input state for subsequent measurement cycles and matching the expansion rates. Second order gravity gradients generated by the payload itself~\cite{DAgostino2011Met} can introduce a substantial bias, several orders of magnitude above the estimation for Earth's contribution. Here, a significant advantage of the satellite mission compared to Earth-based experiments is exploited: Given this bias is sufficiently stable in time, it can be removed due to the calibration measurement at apogee where a gravity dependent violation signal would vanish. Residual mean field energy of the atomic ensembles requires a careful tuning of the relative atom number.
 \begin{table}[tb]
 \caption{Estimated error budget for the dual species atom interferometer with $^{87}$Rb and $^{85}$Rb. The uncertainty in differential acceleration of $7.9\cdot 10^{-15}\,\mathrm{ms}^{-2}$ is evaluated at perigee for an altitude of $700\,\mathrm{km}$. Uncertainties due to the overlap are treated as correlated as are the two contributions due to the effective wave front curvature. All other terms are expected to be uncorrelated.}
 \begin{tabular}{p{0.22\textwidth}p{0.29\textwidth}cp{0.24\textwidth}}\hline
   \textbf{Error source} & \textbf{Conditions} & \textbf{Limit} & \textbf{Comment} \\
   &  & \textbf{in $10^{-15}\,\mathrm{ms}^{-2}$} & \\ \hline
  Gravity gradient & $\Delta z=1.1\cdot10^{-9}\,\mathrm{m}$ & $2.5$ & Connected to\\
  & $\Delta v_{\mathrm{z}}=3.1\cdot10^{-10}\,\mathrm{m}$ & $3.5$ & magnetic field\\
  Coriolis & $\Delta v_{\mathrm{x}}=3.1\cdot10^{-10}\,\mathrm{m}$ & $6.2\cdot10^{-1}$ & gradient\\
  acceleration & $\Delta v_{\mathrm{y}}=3.1\cdot10^{-10}\,\mathrm{m}$ & $6.2\cdot10^{-1}$ & and distance\\
  Other terms & $\Delta x=1.1\cdot10^{-9}\,\mathrm{m}$ & $5.5\cdot10^{-2}$ & to center of mass\\
  depending & $\Delta y=1.1\cdot10^{-9}\,\mathrm{m}$ & $1.6\cdot10^{-3}$ & \\
  on the overlap & others & $4.6\cdot10^{-2}$ & \\ \hline
  Photon recoil & $T_{\mathrm{zzz}}=6GM_{\mathrm{e}}/R^{4}$ & $3.9\cdot10^{-2}$ & \\ \hline
  Self-gravity & Null measurement & $1$ & See section~\ref{sec:orbit} \\ \hline
  Static magnetic & $B_{0}=1\,\mathrm{mG}$, & $1$ & Relieved by input \\ 
  fields & $\nabla B_{0}=1\,$\textmu$\mathrm{G/m}$ &  & state reversal \\ \hline
  Effective wave  & Mirror curvature & $6.3\cdot10^{-1}$ & Relaxed by \\
  front curvature & $R=250\,\mathrm{km}$, &  & expansion rate \\
  & initial coll. to $\approx400\,\mathrm{m}$, $T_{\mathrm{at}}\approx0.07\,\mathrm{nK}$ & $2.8\cdot10^{-1}$ & match \\ \hline
  Mean field & Beam splitter & $2$ & \\ 
   & accuracy $0.1\,\%$, &  &  \\   
   & $N_{87}\approx1.697\left(\pm 0.001\right)N_{85}$ &  &  \\ \hline
  Spurious & Spurious acceleration & $1$ & \\
  {accelerations} & $4\cdot10^{-7}\,\mathrm{ms}^{-2}$, suppression ratio $2.5\cdot10^{-9}$ &  &  \\ \hline
  Detection & $|\epsilon-1|<0.003$ & $<1$ & Post correction \\
  {efficiency} &  &  & from Bayesian fit \\ \hline
  Sum &  & $7.9$ & \\ \hline
 \end{tabular}
 \label{tab:error_budget_table}
 \end{table}
\section{Conclusion}
In this work, we presented an interferometer scheme for a space borne test of the UFF with a dual species $^{87}$Rb / $^{85}$Rb atom interferometer in the scope of the STE-QUEST~\cite{missionpaper2013} mission. Implemented mitigation techniques were discussed and the resulting performance and error budget estimated. The two atomic species are simultaneously interrogated by symmetric Raman beam splitters forming a Mach-Zehnder geometry with a free evolution time of $5\,\mathrm{s}$ enabled by the zero-g environment. Expansion rates of $82\,$\textmu$\mathrm{m/s}$ ensure a high contrast despite the velocity selectivity of the beam splitting process and velocity dependent dephasing. The specific choice of $^{87}$Rb / $^{85}$Rb allows the engineering of a $2.5\cdot10^{-9}$ suppression ratio for linear accelerations and vibrations as well as matching the expansion rates to compensate wave front curvature induced bias terms. To reduce the impact of magnetic field gradients a reversal of the internal interferometer input states is foreseen. Based on this approach, the effects of environmental parameters as spurious accelerations, rotations, and gravity gradients as well as payload specific requirements as the overlap of the two ensembles, magnetic field gradients, and wave front curvature are evaluated. Noise sources are assessed as compatible with a shot noise limited sensitivity for differential accelerations of $3\cdot10^{-12}\,\mathrm{ms}^{-2}$ per cycle of $20\,\mathrm{s}$. In the STE-QUEST reference orbit with an inertial pointing spacecraft, a sensitivity to the E\"otv\"os ratio of better than $5.3\cdot10^{-14}$ per revolution can be obtained, implying an integrated sensitivity better than $2\cdot10^{-15}$ after $1.5\,\mathrm{years}$. Uncertainties in the bias terms are estimated to be compatible with the targeted inaccuracy of $2\cdot10^{-15}$.\\
The present analysis is performed for the specific STE-QUEST mission but be can easily be extended to different platforms as e.g. circular orbits, nadir pointing modes, or to other environmental constraints, different parameter sets for the atom interferometer. On the contrary, direct extrapolation for Earth-based experiments is difficult. Indeed, most of the systematic effects detailed in table~\ref{tab:error_budget_table} are larger by orders of magnitudes. Firstly, most of them are connected to the ability of superimposing the centres of mass of the two atomic species, which suffers from a large sag in the trap position under gravity. Secondly, some of the major systematics depend directly on the global displacement of the atomic wave packets compared to the payload itself, which is $12\,\mathrm{cm}$ for $2T=10\,\mathrm{s}$ and double diffraction in space and $15\,\mathrm{m}$ on ground for $2T=2\,\mathrm{s}$ and a $50\,\mathrm{th}$ order Bragg transition to reach the same scaling factor. Finally, a space experiment can take the advantage of calibration residual bias thanks to the possibility of modulating the signal of interest, by for example retuning the direction of the satellite compared to the direction of gravity. Although no space borne atom interferometry experiments exist yet, several national activities push scientific investigations and technology developments in zero-g environments as parabolic flights~\cite{Geiger2011NatComm,Nyman2006APB} and drop tower experiments~\cite{Muntinga2013PRL,Rudolph2011MST,Zoest2010Science}.
\ack
This work was supported by the German Space Agency Deutsches Zentrum f\"ur Luft- und Raumfahrt with funds provided by the Federal Ministry of Economics and Technology under grant numbers 50~OY~1303 and 50 OY 1304, the German Research
Foundation (DFG) by funding the Cluster of Excellence
QUEST Centre for Quantum Engineering and
Space-Time Research, the French space agency Centre National d'Etudes Spatiales, and the European Space Agency.

\section*{References}
\bibliography{STE-QUEST_error_assessment}

\end{document}